\documentstyle[aps,prd,twocolumn,amsfonts,epsfig,floats]{revtex}

\newcommand{\mb}[1]{\mbox{\boldmath $#1$}}

\begin{document} 
\draft \twocolumn[\hsize\textwidth\columnwidth\hsize\csname
@twocolumnfalse\endcsname

\title{Bulk effects in the cosmological dynamics of brane-world scenarios}

\author{Antonio Campos and Carlos F. Sopuerta}

\address{~}

\address{Relativity and Cosmology Group, School of Computer Science
and Mathematics, \\
Portsmouth University, Portsmouth~PO1~2EG, United Kingdom}

\address{~}

\date{\today}

\maketitle 


\begin{abstract}
In this paper we deal with the cosmological dynamics of Randall-Sundrum 
brane-world type scenarios in which the five-dimensional Weyl tensor has 
a non-vanishing projection onto the three-brane where matter fields are 
confined.  Using dynamical systems techniques, we study how the state 
space of Friedmann-Lema\^{\i}tre-Robertson-Walker (FLRW) and Bianchi 
type I cosmological models is affected by the bulk Weyl tensor, focusing
in the differences that appear with respect to general relativity
and also Randall-Sundrum cosmological scenarios without the Weyl tensor 
contribution.
\end{abstract}

\vskip 1pc  ] 



\section{Introduction}\label{sec1}

Recently, Randall and Sundrum have shown that for non-factorizable 
geometries in five dimensions the zero-mode of the Kaluza-Klein 
dimensional reduction can be localized in a four-dimensional submanifold 
where Newtonian gravity is effectively reproduced at large 
distances~\cite{RanSun:1999b}.  
The picture coming out is one with all matter and gauge fields,
excepting gravity, confined in a three-brane embedded in a five-dimensional 
spacetime with ${\mathbb{Z}}_2$ symmetry.
This scenario was motivated by orbifold compactification of higher
dimensional string theories.  In particular by the dimensional reduction 
of eleven-dimensional supergravity in 
${\mathbb{R}}^{10}\times S^1/{\mathbb{Z}}_2$ introduced by 
Ho$\check{\mbox{r}}$ava and Witten\cite{HorWit:1996}.
The cosmological implications of the Ho$\check{\mbox{r}}$ava-Witten 
theory have been analyzed in~\cite{HOWI}.  The actual relevance of
this type of scenarios is that they provide an alternative framework
to investigate new solutions to the hierarchy and cosmological constant
problems.

Gravity on the brane can be described by the Einstein equations modified
by two additional terms, one quadratic in the matter variables and
the second one being the ``electric'' part of the five-dimensional Weyl 
tensor~\cite{BinDefLan:2000,BinDefEllLan:2000,ShiMaeSas:2000,SasShiMae:2000}.  
In a previous work~\cite{CamSop:2001} we studied 
the whole dynamics of the Friedmann-Lema\^{\i}tre-Robertson-Walker 
(FLRW) and the Bianchi type I and V cosmological models taking into account 
only the effects produced by the first type of corrections.  
In that paper we found new critical points corresponding to the 
Bin$\acute{\mbox{e}}$truy-Deffayet-Langlois (BDL) 
models~\cite{BinDefLan:2000}, representing 
the dynamics at very high energies, where the extra-dimension 
effects become dominant.  The presence of these models seems to be a 
generic feature of the state space of more general cosmological models. 
Moreover, we saw that the state space presents new bifurcations
and how the dynamical character of some of the critical points
changes with respect to the general-relativistic case.
Finally, we showed that for models satisfying all the ordinary energy 
conditions and causality requirements the anisotropy is negligible near 
the initial singularity. On the other hand, for late times the anisotropy 
is diluted as in general relativity (GR) but now there are intermediate 
stages in which the shear contribution grows.

Here, we will complete the analysis done in~\cite{CamSop:2001} 
(hereafter Paper I) studying the effects of the five-dimensional 
Weyl tensor on FLRW and Bianchi type I cosmological models.  In the case 
of FLRW models this study will be completely general whereas in the 
Bianchi type I case we will neglect the Weyl tensor components for 
which the theory does not provide evolution equations.  The paper is 
organized as follows.   In Sec.~\ref{prel} we briefly describe the 
geometric formulation of brane-world scenarios and the dynamical 
equations for FLRW and Bianchi type I cosmological models with a 
perfect-fluid type matter content.  The dynamics of the FLRW 
and Bianchi type I cosmological models will be analyzed in 
Secs.~\ref{sec3} and~\ref{sec4} respectively.
We will finish with some concluding remarks in Sec.~\ref{core}.
Through this paper we will follow the following notation:
upper-case Latin letters denote coordinate indices in the bulk spacetime 
($A,B,\ldots = 0 - 4$) whereas lower-case Latin letters denote coordinate 
indices in the brane ($a,b,\ldots=0 - 3$).  We use physical units in 
which $c=1$.



\section{Preliminaries}\label{prel}
In what follows we introduce the geometric framework to analyze
the brane-world scenario and the main assumptions we will adopt to
study the FLRW and Bianchi type I cosmological models.

\subsection{Geometric formulation of brane-world scenarios}

In Randall-Sundrum brane-world type scenarios matter fields are
confined in a three-brane embedded in a five-dimensional spacetime (bulk).
It is assumed that the metric of this spacetime,
 $g^{(5)}_{AB}$, obeys the 
Einstein equations with a negative cosmological constant
$\Lambda_{(5)}$ 
(see~\cite{ShiMaeSas:2000,SasShiMae:2000,Maa:2000})\begin{equation}
G^{(5)}_{AB} = -\Lambda_{(5)}g^{(5)}_{AB}
+\kappa^2_{(5)} \delta(\chi)\left[ -\lambda\,g_{AB}+T_{AB}
\right]\,,  \label{eeob}
\end{equation}
where $G^{(5)}_{AB}$ denotes the Einstein tensor and $\kappa_{(5)}$ is 
the five-dimensional gravitational coupling constant. Moreover, $T_{AB}$ 
is the matter energy-momentum tensor and the Dirac delta function reflects 
the fact that matter is confined in the spacelike hypersurface 
$x^4\equiv\chi=0$, the three-brane, with induced metric $g_{AB}$ and tension 
$\lambda\,.$   

From the Gauss-Codacci equations, the Israel junction conditions and
the ${\mathbb{Z}}_2$ symmetry with respect to the three-brane, we can 
see~\cite{ShiMaeSas:2000,SasShiMae:2000} that the effective 
Einstein equations on the brane are
\begin{equation}
G_{ab}=-\Lambda g_{ab}+\kappa^2 T_{ab}+
\kappa^4_{(5)} S_{ab} - E^{(5)}_{ab} \,, \label{mefe}
\end{equation}
where $G_{ab}$ is the Einstein tensor of the induced metric $g_{ab}\,.$ 
The four-dimensional gravitational constant 
$\kappa$ and the cosmological constant $\Lambda$ are given in terms 
of the fundamental constants in the bulk $(\kappa_{(5)},\Lambda_{(5)})$ and 
the brane tension $\lambda$.  It is important to note that in order to 
recover conventional gravity on the brane $\lambda$ must be assumed to 
be positive.
As one can observe, there are two corrections to
the general-relativistic equations.  The first one, denoted by the
term $S_{ab}$, is quadratic in the matter energy-momentum tensor
$T_{ab}$ ($T\equiv g^{ab}T_{ab}$)
\begin{equation}
S_{ab} = \textstyle{1\over12}T T_{ab}-\textstyle{1\over4}
T_a{}^c T_{bc}+\textstyle{1\over24}g_{ab}\left[3T^{cd}
T_{cd}-T^2\right] \,. \label{cobu}
\end{equation}
The second correction, $E^{(5)}_{ab}$, correspond to the 
``electric'' part of the five-dimensional Weyl tensor, 
$C^{(5)}_{ABCD}$, with respect to the normal, $n_A$ ($n^An_A=1$), to
the hypersurface $\chi=0$, that is
\[ E^{(5)}_{AB} = C^{(5)}_{ACBD}n^Cn^D \,. \]
From the modified Einstein equations~(\ref{mefe}) and the energy-momentum
tensor conservation equations ($\nabla_a T^a{}_b=0$) we get 
a constraint on $S_{ab}$ and $E^{(5)}_{ab}$
\begin{equation} 
\nabla^a (E^{(5)}_{ab}-\kappa^4_{(5)}S_{ab}) = 0 \,. \label{coco}
\end{equation}

Following~\cite{Maa:2000}, it is useful to decompose $E^{(5)}_{ab}$
with respect to any timelike observers $\mb{u}$ ($u^au_a=-1$) in a
scalar part, 
${\cal U}$, a vector part, ${\cal Q}_a$, and a tensorial part 
${\cal P}_{ab}$, in such a way that $E^{(5)}_{ab}$ can be
written as 
\begin{equation} 
E^{(5)}_{ab} = -\left(\frac{\kappa^{}_{(5)}}{\kappa}\right)^4
\left[ (u_au_b+\textstyle{1\over3}h_{ab}){\cal U} +2u_{(a}{\cal Q}_{b)}
+{\cal P}_{ab}\right]\,, \label{spli}
\end{equation}
where the following properties hold
\[ {\cal Q}_au^a=0\,,~~~ {\cal P}_{(ab)}= {\cal P}_{ab}\,,~~
{\cal P}^a{}_a=0\,,~~ {\cal P}_{ab}u^b=0 \,. \]
That is, ${\cal Q}_a$ is a spatial vector and ${\cal P}_{ab}$ is a
spatial, symmetric and trace-free tensor.  The scalar term has the
same form as the energy-momentum tensor of a radiation perfect fluid,
and for this reason ${\cal U}$ was called ``dark'' energy density
in~\cite{Maa:2000}.  Note that the 
constraint~(\ref{coco})
provides evolution equations for ${\cal U}$ and ${\cal Q}_a$, but
not for ${\cal P}_{ab}$ (see~\cite{Maa:2000}).

\subsection{Cosmological dynamics on the brane}
Throughout this paper we will describe matter by a perfect fluid 
\[ T_{ab} = (\rho+p)u_au_b + pg_{ab} \,, \]
where $\mb{u}$, $\rho$ and $p$ are the unit fluid velocity of matter, 
the energy density and the pressure of the matter fluid respectively.   
We will also assume a linear barotropic equation of state for the fluid 
\begin{equation} 
p = (\gamma-1)\rho\,. \label{leoe}
\end{equation}
The weak energy condition (see, e.g.,~\cite{HawEll:1973}) imposes the
restriction $\rho\geq 0$ and from causality requirements we have that 
the parameter $\gamma\in[0,2]$.

In this paper we deal with non-tilted homogeneous cosmological models on
the brane, i.e. we are assuming that the fluid velocity is orthogonal to 
the hypersurfaces of homogeneity.  Then, it is convenient to make the 
splitting~(\ref{spli}) with respect to the fluid velocity.  
In particular we will consider: (i) The FLRW models, with metric tensor
given by
\[ \mbox{ds}^2 = -dt^2 + a^2(t)\left[ dr^2+\Sigma^2_k(r)(
 d\theta^2+\sin^2\theta d\varphi^2)\right] \,, \] 
where $a(t)$ is the scale factor, $k$ is the curvature parameter, and
\[ \Sigma_k(r) = \left\{ \begin{array}{ll}
\sin r & \mbox{for $k=1$}\,, \\
r & \mbox{for $k=0$}\,, \\
\sinh r & \mbox{for $k=-1$}\,; \end{array} \right. \]
and (ii) Bianchi type I models, whose line element can be written as
\begin{equation} 
\mbox{ds}^2 = -dt^2 + \sum_{\alpha=1}^3 A^2_\alpha(t)(dx^\alpha)^2\,. 
\label{lebi}
\end{equation}
Then, in both cases the fluid velocity is $\vec{\mb{u}}=\mb{\partial}
/\mb{\partial t}\,.$  We will not consider Bianchi type V models as we did
in Paper I because, as we showed there, their main dynamical features are
a combination of those of the FLRW and Bianchi type I models.

Taking into account these assumptions and the effective Einstein's 
equations~(\ref{mefe}), the consequences of having a
FLRW model on the brane are
\[ {\cal Q}_a = {\cal P}_{ab} = 0 \,, \]
and this, through the constraint~(\ref{coco}), further implies
\begin{equation}  
D_a{\cal U}=0 ~ \Leftrightarrow ~ {\cal U}={\cal U}(t)\,,\label{grau}  
\end{equation}
where $D_a$ denotes the covariant derivative associated with the
induced metric on the hypersurfaces of homogeneity ($h_{ab}\equiv
g_{ab}+u_au_b$).  Instead, in the case of Bianchi type I we get 
${\cal Q}_a=0$ but we do not get any restriction on 
${\cal P}_{ab}\,.$  Since there is no way of fixing the dynamics
of this tensor we will study the particular case in which it is zero.  
Then, also the condition~(\ref{grau}) follows.

The key equation to describe the dynamics of these models is the 
Friedmann equation, which determines the expansion of the universe,
or in other words, the Hubble function: $3H \equiv \nabla_a u^a\,.$ 
The general expression of this equation in the case of homogeneous
cosmological models is
\begin{equation} 
H^2 = \frac{1}{3}\kappa^2\rho\left(1+\frac{\rho}{2\lambda}\right)
-\frac{1}{6}{}^3R+\frac{1}{3}\sigma^2+\frac{1}{3}\Lambda +
\frac{2{\cal U}}{\lambda\kappa^2}\,, \label{frie}
\end{equation}
where ${}^3R$ is the scalar curvature of the hypersurfaces orthogonal
to the fluid flow and $2\sigma^2\equiv\sigma^{ab}\sigma_{ab}$ is the
shear scalar ($\sigma_{ab}\equiv h_a{}^ch_b{}^d\nabla_{(c}u_{d)}-
Hh_{ab}$).  We consider only the case of a positive fourd-dimensional
cosmological constant, i.e. $\Lambda\geq 0$.   
For Bianchi type I models ${}^3R$ vanishes whereas for FLRW
models it is given by ${}^3R=6ka^{-2}(t)\,.$ On the other hand,
the shear vanishes for FLRW models and for Bianchi type I models 
the evolution of the shear scalar is
\[ (\sigma^2)^\cdot = -6H\sigma^2 \,. \]
In both cases the evolution equation for ${\cal U}$ is~\cite{Maa:2000}
\[ \dot{\cal U} = - 4H{\cal U}\,, \]
and, from the energy-momentum conservation equation, the evolution
of $\rho$ is simply
\begin{equation} 
\dot{\rho}= -3\gamma H\rho \,. \label{emce}
\end{equation}
Finally, an important quantity in the dynamical study will be the 
deceleration parameter $q$, defined by
\begin{equation} 
\dot{H} \equiv -(1+q)H^2 \,, \label{decd}
\end{equation}
where $\dot{H}$ is found to be
\begin{eqnarray}
\dot{H} & = & -H^2-\frac{3\gamma-2}{6}\kappa^2\rho\left[1+\frac{3\gamma-1}
{3\gamma-2}\frac{\rho}{\lambda}\right]-\frac{2}{3}\sigma^2 \nonumber \\
& & +\frac{1}{3}\Lambda-\frac{2{\cal U}}{\lambda\kappa^2}
\,,\label{raye}
\end{eqnarray}
which is the modified Raychaudhuri equation.



\section{Dynamics of the FLRW models}\label{sec3} 
In Paper I we studied the dynamics of the FLRW cosmological models 
in a brane-world scenario in which the effects of the extra dimensions
are only due to the term quadratic in the matter sources, $S_{ab}$ 
[see Eq.~(\ref{mefe})].  Now, we will add the contribution of
$E^{(5)}_{ab}\,,$ which, as we have seen above, only contains the
scalar part ${\cal U}\,.$  
In order to study the dynamics of these models we will closely follow 
the analysis carried out in Paper I.  This procedure is based on the 
introduction of new variables which describe the dynamics in such
a way that the state space is a compact space (see~\cite{DSCO,DSTH} 
for details on the techniques of dynamical systems analysis).  
To carry out this study we have to consider four differentiated cases 
according to the signs of 
${}^3R$ and ${\cal U}$:  (A) ${\cal U}\geq 0$ and ${}^3R\leq 0$;  
(B) ${\cal U}\geq 0$ and ${}^3R\geq 0$; (C) ${\cal U}\leq 0$ and 
${}^3R\leq 0$; (D) ${\cal U}\leq 0$ and ${}^3R\geq 0$.

\subsection{Case ${\cal U}\geq 0$ and ${}^3R\leq 0$}\label{frwa}
The state space in this case can be described by the 
dimensionless general-relativistic variables
\begin{equation}
\Omega_\rho\equiv \frac{\kappa^2\rho}{3H^2}\,,   ~~ 
\Omega_k\equiv -\frac{{}^3R}{6H^2}\,,  ~~ 
\Omega_\Lambda\equiv \frac{\Lambda}{3H^2}\,, \label{dlv1} 
\end{equation}
and the following new variables
\begin{equation}
\Omega_\lambda\equiv \frac{\kappa^2\rho^2}{6\lambda H^2} 
\,,  ~~ \Omega_{\cal U} \equiv  \frac{2{\cal U}}{\lambda\kappa^2H^2} 
\,. \label{dlv2}
\end{equation}
Here $\Omega_\rho$ is the ordinary density parameter and 
$\Omega_k$, $\Omega_\Lambda$, $\Omega_\lambda$, and $\Omega_{\cal U}$
are the fractional contributions of the curvature, cosmological constant, 
brane tension, and the dark radiation energy, respectively, to the universe 
expansion~(\ref{frie}).  In fact, the Friedmann equation~(\ref{frie}) 
becomes
\[ \Omega_\rho+\Omega_k+\Omega_\Lambda+\Omega_\lambda+\Omega_{\cal U} =1\,.\]
Since these terms are non-negative they must belong to the interval
$[0,1]$ and hence, the variables $\mb{\Omega}=(\Omega_\rho,\Omega_k,
\Omega_\Lambda,\Omega_\lambda,\Omega_{\cal U})$ define a compact state
space.  Introducing the time derivative $|H|^{-1}d/dt$, which we will
denote by a prime, the evolution equation for $H$ [Eq.~(\ref{decd})]
will decouple from the equations for $\mb{\Omega}$. Then, the
set of equations describing the dynamics is given by
\begin{eqnarray}
\Omega'_\rho & = & \epsilon [2(1+q)-3\gamma]\Omega_\rho \,,
\label{c1i} \\
\Omega'_k & = & 2\epsilon q\Omega_k \,, \\
\Omega'_\Lambda & = & 2\epsilon(1+q)\Omega_\Lambda \,, \\
\Omega'_\lambda & = & 2\epsilon (1+q-3\gamma)\Omega_\lambda \,, \\
\Omega'_{\cal U} & = & 2\epsilon(q-1)\Omega_{\cal U}\,, \label{c1f}
\end{eqnarray}
where $q$  is
\[ q = \frac{3\gamma-2}{2}\Omega_\rho-\Omega_\Lambda+(3\gamma-1)
\Omega_\lambda+\Omega_{\cal U} \,, \]
and $\epsilon$ is the sign of $H\,.$ As is clear, for $\epsilon=1$ the 
model will be in expansion, and for $\epsilon=-1$ it will be in contraction.

The critical points of the dynamical system~(\ref{c1i})-(\ref{c1f}),
their coordinates in the state space and their eigenvalues are given in the
following table~\cite{Nota}
\begin{quasitable}
\begin{tabular}{ccc}
Model  & Coordinates   & Eigenvalues  \\ \tableline
$\mbox{F}_\epsilon$ & $(1,0,0,0,0)$ & $\epsilon(3\gamma-2,3\gamma-2,
3\gamma,-3\gamma,3\gamma-4)$ \\
$\mbox{M}_\epsilon$ & $(0,1,0,0,0)$ & $\epsilon(-(3\gamma-2),0,2,
-2(3\gamma-1),-2)$ \\
$\mbox{dS}_\epsilon$ & $(0,0,1,0,0)$ & $-\epsilon(3\gamma,2,2,6\gamma,4)$  \\
$\mbox{m}_\epsilon$ & $(0,0,0,1,0)$ & $2\epsilon(\textstyle{3\gamma\over2},
(3\gamma-1),3\gamma,3\gamma-1,3\gamma-2)$ \\
$\mbox{R}_\epsilon$ & $(0,0,0,0,1)$ & $\epsilon(-(3\gamma-4),2,4,-2(3\gamma-2),
2)$
\end{tabular}
\end{quasitable}
We have five hyperbolic critical points: 
the flat FLRW models ($\mbox{F}$), $k=\Lambda=\lambda^{-1}={\cal U}=0$
and $a(t)=t^{2/(3\gamma)}$;
the Milne universe ($\mbox{M}$), $\rho=\Lambda={\cal U}=0\,,$ $k=-1$ and 
$a(t)=t$; the de Sitter model ($\mbox{dS}$), $k=\rho={\cal U}=0$ and 
$a(t)=\exp(\sqrt{\Lambda/3}\,t)$; the non-general-relativistic 
BDL model 
($\mbox{m}$)~\cite{BinDefLan:2000,BinDefEllLan:2000,FlaTyeWas:2000b}, 
found through a limiting process to be $a(t)=t^{1/(3\gamma)}$ (see 
Paper I for more details);
and a model ($\mbox{R}$) with the metric of a flat radiation FLRW model, 
$a(t)=t^{1/2}$, but characterized by $\rho=k=\Lambda=\lambda^{-1}=0$.

\subsection{Case ${\cal U}\geq 0$ and ${}^3R\geq 0$}\label{frwb}
Here we can follow the study corresponding to the non-negative scalar
curvature case in Paper I.  Then, 
we will consider the variables $\mb{\tilde{\Omega}}\equiv(Q,
\tilde{\Omega}_\rho,\tilde{\Omega}_\Lambda,\tilde{\Omega}_\lambda,
\tilde{\Omega}_{\cal U})$, where Q is 
\[ Q\equiv\frac{H}{D}\,,~~~ D^2\equiv H^2+\textstyle{1\over6}{}^3R\,. \]
and the variables with tilde are the analogues of those in 
(\ref{dlv1}) and (\ref{dlv2}) but normalized with respect to $D$ instead of 
$H$.  Then, the Friedmann equation~(\ref{frie}) is now
\begin{equation}
\tilde{\Omega}_\rho+\tilde{\Omega}_\Lambda+\tilde{\Omega}_\lambda
+\tilde{\Omega}_{\cal U} = 1 \,. \label{frib}
\end{equation}
Taking into account that $q$ is now given by
\[ 1+qQ^2=\textstyle{{3\gamma}\over2}(\tilde{\Omega}_\rho+
2\tilde{\Omega}_\lambda)+2\tilde{\Omega}_{\cal U} \,, \]
and using the time derivative $'\equiv D^{-1}d/dt\,,$
the dynamical equations for $\mb{\tilde{\Omega}}$ are decoupled from 
$D$ and are given by
\begin{eqnarray*}
Q' & = & -qQ^2(1-Q^2)\,, \\
\tilde{\Omega}'_\rho & = & [2(1+qQ^2)-3\gamma]Q\tilde{\Omega}_\rho \,, \\
\tilde{\Omega}'_\Lambda & = & 2(1+qQ^2)Q\tilde{\Omega}_\Lambda \,, \\
\tilde{\Omega}'_\lambda & = & 2[1+qQ^2-3\gamma]Q\tilde{\Omega}_\lambda\,, \\
\tilde{\Omega}'_{\cal U} & = & -2(1-qQ^2)Q\tilde{\Omega}_{\cal U} \,.
\end{eqnarray*}
The critical points of this dynamical system, as well as their coordinates 
and eigenvalues are~\cite{Nota}
\begin{quasitable}
\begin{tabular}{ccc}
Model  & Coordinates   & Eigenvalues \\ \tableline
$\mbox{F}_\epsilon$ & $(\epsilon,1,0,0,0)$ & $\epsilon(3\gamma-2,3\gamma,
3\gamma,-3\gamma,3\gamma-4)$ \\
$\mbox{dS}_\epsilon$ & $(\epsilon,0,1,0,0)$ & $-\epsilon(2,3\gamma,0,6\gamma,4)$ \\
$\mbox{E}~$ & $(0,\tilde{\Omega}^\ast_\rho,\tilde{\Omega}^\ast_\Lambda,
\tilde{\Omega}^\ast_\lambda,\tilde{\Omega}^\ast_{\cal U})$ &
$(0,\sqrt{\phi},0,-\sqrt{\phi},0)$ \\
$\mbox{m}_\epsilon$ & $(\epsilon,0,0,1,0)$ & $2\epsilon(3\gamma-1,
\textstyle{{3\gamma}\over2},3\gamma,3\gamma,3\gamma-2)$ \\
$\mbox{R}_\epsilon$ & $(\epsilon,0,0,0,1)$ & $\epsilon(2,-(3\gamma-4),4,
-2(3\gamma-2),4)$
\end{tabular}
\end{quasitable}
Here $\mbox{E}$ denotes a infinite set of saddle points whose line
element is that of the Einstein universe ($k=1$ and $H=0$). 
Their coordinates $\mb{\tilde{\Omega}^\ast}$ satisfy~(\ref{frib}) 
and
\begin{equation} 
1-\frac{3\gamma}{2}(\tilde{\Omega}^\ast_\rho+2\tilde{\Omega}^\ast_\lambda)
-2\tilde{\Omega}^\ast_{\cal U}=0 \,. \label{coiv}
\end{equation}
In the table above $\phi$ is given by 
\[ \phi = \frac{3\gamma}{2}\left[ \left(3\gamma-2\right)
\tilde{\Omega}^\ast_\rho+4\left(3\gamma-1\right)
\tilde{\Omega}^\ast_\lambda\right]+4\tilde{\Omega}^\ast_{\cal U}\,. \]
From the relation~(\ref{coiv}) and the allowed values of the 
variables $\mb{\tilde{\Omega}}$ one can show that $\phi$ is always
positive for any $\gamma\,.$  
In GR the Einstein universe appears for $\gamma \geq 2/3\,.$
In Paper I we showed that in brane-world scenarios without the contribution
coming from the five-dimensional Weyl tensor it appears for $\gamma \geq
1/3\,.$ Now, considering also this contribution Eq.~(\ref{coiv}) shows that 
the Einstein universe appears for any value of $\gamma\,.$

\subsection{Case ${\cal U}\leq 0$ and ${}^3R\leq 0$}\label{frwc}
In this case we can get a compact state space by introducing the following
set of dynamical variables
$\mb{\bar{\Omega}}\equiv(Z,\bar{\Omega}_\rho,\bar{\Omega}_k,
\bar{\Omega}_\Lambda,\bar{\Omega}_\lambda)$, where
\begin{equation} 
Z\equiv\frac{H}{N}\,,~~N^2 \equiv H^2
-\frac{2{\cal U}}{\lambda\kappa^2}\,,\label{deom}
\end{equation}
and the variables with a bar are like those in~(\ref{dlv1}) and~(\ref{dlv2})
but normalized with respect to $N$ instead of $H$.  As before,
the Friedmann equation~(\ref{frie}) reduces to an expression in which 
all the terms are non-negative 
\begin{equation}
\bar{\Omega}_\rho+\bar{\Omega}_k+\bar{\Omega}_\Lambda+
\bar{\Omega}_\lambda = 1\,, \label{frnn}
\end{equation}
whence it follows that all of them are smaller than $1\,.$
Using the time derivative $'\equiv N^{-1}d/dt\,,$ in order to 
decouple the evolution of $N$ from the rest of variables, and taking into
account the expression for $q$
\[ (q-1)Z^2 = \textstyle{{3\gamma}\over2}(\bar{\Omega}_\rho+
2\bar{\Omega}_\lambda)+\bar{\Omega}_k-2\,, \]
the dynamical system for $\mb{\bar{\Omega}}$ is 
\begin{eqnarray*}
Z' & = & -(q-1)Z^2(1-Z^2)\,, \\
\bar{\Omega}'_\rho & = & -[(3\gamma-4)-2(q-1)Z^2]Z\bar{\Omega}_\rho \,, \\
\bar{\Omega}'_k & = & 2[1+(q-1)Z^2]Z\bar{\Omega}_k\,, \\
\bar{\Omega}'_\Lambda & = & 2[2+(q-1)Z^2]Z\bar{\Omega}_\Lambda\,, \\
\bar{\Omega}'_\lambda & = & -2[(3\gamma-2)-(q-1)Z^2]Z\bar{\Omega}_\lambda\,.
\end{eqnarray*}
The critical points, their coordinates and eigenvalues are given in the 
table bellow~\cite{Nota}
\begin{quasitable}
\begin{tabular}{ccc}
Model  & Coordinates   & Eigenvalues \\ \tableline
$\mbox{F}_\epsilon$ & $(\epsilon,1,0,0,0)$ & $\epsilon(3\gamma-4,3\gamma,
3\gamma-2,3\gamma,-3\gamma)$ \\
$\mbox{M}_\epsilon$ & $(\epsilon,0,1,0,0)$ & $\epsilon(-2,-(3\gamma-2),2,
2,-2(3\gamma-1))$ \\
$\mbox{dS}_\epsilon$ & $(\epsilon,0,0,1,0)$ & $-\epsilon(4,3\gamma,2,0,
6\gamma)$ \\
$\mbox{m}_\epsilon$ & $(\epsilon,0,0,0,1)$ & $2\epsilon(3\gamma-2,
\textstyle{3\gamma\over2},3\gamma-1,3\gamma,3\gamma)$ \\
$\mbox{S}~$ & $(0,\bar{\Omega}^\ast_\rho,\bar{\Omega}^\ast_k,
\bar{\Omega}^\ast_\Lambda,\bar{\Omega}^\ast_\lambda)$ &
$(0,\sqrt{\psi},0,0,-\sqrt{\psi})$ \\
\end{tabular}
\end{quasitable}
We find new critical points, denoted by $\mbox{S}$, representing static
models with vanishing and negative spatial curvature, i.e. $H=0$ and 
$k=0,-1\,.$  Moreover
$\bar{\Omega}^\ast_\rho\,,$ $\bar{\Omega}^\ast_k\,,$
$\bar{\Omega}^\ast_\Lambda\,,$ and $\bar{\Omega}^\ast_\lambda$ are real
numbers satisfying~(\ref{frnn}) and the following relation
\begin{equation} 
2-\bar{\Omega}^\ast_k-\frac{3\gamma}{2}(\bar{\Omega}^\ast_\rho+
2\bar{\Omega}^\ast_\lambda)=0 \,. \label{sdef}
\end{equation}
The eigenvalues of $\mbox{S}$ are given in terms of $\psi$: 
\begin{eqnarray} 
\psi = \frac{3\gamma}{2}\left[(3\gamma-4)\bar{\Omega}^\ast_\rho+
4(3\gamma-2)\bar{\Omega}^\ast_\lambda\right]-2\bar{\Omega}^\ast_k \,.  
\label{psie}
\end{eqnarray}
Eq.~(\ref{sdef}) implies that these static models ($\mbox{S}$) only 
appear for $\gamma\geq 2/3$.  Furthermore $a\,,$ $\rho$ and ${\cal U}$ 
are constants, $(a^\ast,\rho^\ast,{\cal U}^\ast)$, which must
satisfy the following condition 
\begin{equation} 
-{\cal U}^\ast \geq \frac{\lambda\kappa^2}{2}\left(\frac{\Lambda}{3}-
\frac{k}{a^{\ast^2}}\right) \,, \label{cped}
\end{equation}
in order to have a positive energy density.

\subsection{Case ${\cal U}\leq 0$ and ${}^3R\geq 0$}\label{frwd}
This case is more involve since the Friedmann equation~(\ref{frie})
has now two non-positive terms.  Taking this into account we will
consider the following dimensionless dynamical variables $\mb{\hat{\Omega}}=
(W,\hat{\Omega}_\rho,\hat{\Omega}_\Lambda,\hat{\Omega}_\lambda,
\hat{\Omega}_{\cal U})$, where
\[ W\equiv \frac{H}{P}\,,~~~P^2 \equiv H^2+
\frac{1}{6}{}^3R-\frac{2{\cal U}}{\lambda\kappa^2} \,,\]
$\hat{\Omega}_\rho\,,$ $\hat{\Omega}_\Lambda\,,$
$\hat{\Omega}_{\cal U}\,,$ and $\hat{\Omega}_\lambda$
are the analogous of those in~(\ref{dlv1}) and (\ref{dlv2}) but normalized
with $P$.  It is important to note that $\hat{\Omega}_{\cal U}$ does not 
appear in the Friedmann equation~(\ref{frie}), which now reads
\begin{equation}
\hat{\Omega}_\rho+\hat{\Omega}_\Lambda+\hat{\Omega}_\lambda=1\,,
\label{frrr}
\end{equation}
but it will be needed to close the dynamical system.  Moreover, in
this case $\hat{\Omega}_{\cal U}$ is negative and from its definition
it belongs to the interval $[-1,0]\,.$
Then, using the time derivative $' \equiv P^{-1}d/dt$, and the
expression for $q$
\[1+qW^2= \textstyle{{3\gamma}\over2}(\hat{\Omega}_\rho+
2\hat{\Omega}_\lambda)+\hat{\Omega}_{\cal U} \,, \]
the evolution equations for $\mb{\hat{\Omega}}$ are
\begin{eqnarray}
W' & = & -[q(1-W^2)+\hat{\Omega}_{\cal U}]W^2\,, \label{evew} \\
\hat{\Omega}'_\rho & = & [2(1+qW^2)-(2\hat{\Omega}_{\cal U}+3\gamma)
]W\hat{\Omega}_\rho\,, \nonumber \\
\hat{\Omega}'_\Lambda & = & 2(1+qW^2-\hat{\Omega}_{\cal U})
W\hat{\Omega}_\Lambda\,, \nonumber \\
\hat{\Omega}'_\lambda & = & 2[1+qW^2-(\hat{\Omega}_{\cal U}+3\gamma)
]W\hat{\Omega}_\lambda \,, \nonumber \\
\hat{\Omega}'_{\cal U} & = & 2[qW^2-\hat{\Omega}_{\cal U}-1]
W\hat{\Omega}_{\cal U}\,.\nonumber 
\end{eqnarray}
The critical points and their coordinates and eigenvalues are~\cite{Nota}
\begin{quasitable}
\begin{tabular}{ccc}
Model  & Coordinates   & Eigenvalues \\ \tableline
$\mbox{F}_\epsilon$ & $(\epsilon,1,0,0,0)$ & $\epsilon(3\gamma-2,
3\gamma,3\gamma,-3\gamma,3\gamma-4)$ \\
$\mbox{dS}_\epsilon$ & $(\epsilon,0,1,0,0)$ & $-\epsilon(2,3\gamma,0,
6\gamma,4)$ \\
$\mbox{m}_\epsilon$ & $(\epsilon,0,0,1,0)$ & $2\epsilon(3\gamma-2,
\textstyle{{3\gamma}\over2},3\gamma,3\gamma,3\gamma-2)$ \\
$\mbox{E}~$ & $(0,\hat{\Omega}^\ast_\rho,\hat{\Omega}^\ast_\Lambda,
\hat{\Omega}^\ast_\lambda,\hat{\Omega}^\ast_{\cal U})$ &
$(0,\sqrt{\varphi},0,-\sqrt{\varphi},0)$
\end{tabular}
\end{quasitable}
where $\hat{\Omega}^\ast_\rho\,,$ $\hat{\Omega}^\ast_\Lambda\,,$
$\hat{\Omega}^\ast_\lambda\,,$ and $\hat{\Omega}^\ast_{\cal U}$ are real
numbers satisfying (\ref{frrr}) and
\[ \frac{3\gamma}{2}(\hat{\Omega}^\ast_\rho+2\hat{\Omega}^\ast_\lambda)+
\hat{\Omega}^\ast_{\cal U}-1=0\,. \]
Now, the eigenvalues of the Einstein universe are determined by $\varphi$, 
which is given by 
\[ \varphi = \frac{3\gamma}{2}\left[(3\gamma-2)\hat{\Omega}^\ast_\rho+
4(3\gamma-1)\hat{\Omega}^\ast_\lambda\right]+4\hat{\Omega}^\ast_{\cal U}\,.\]
In contrast to what happens in the state space sector analyzed 
in subsection~\ref{frwb} the Einstein universe now appears for
$\gamma\geq 1/3$, just when $\varphi$ starts being non-negative.

\begin{figure*}
\begin{center}
\includegraphics[height=4in,width=6.5in
]{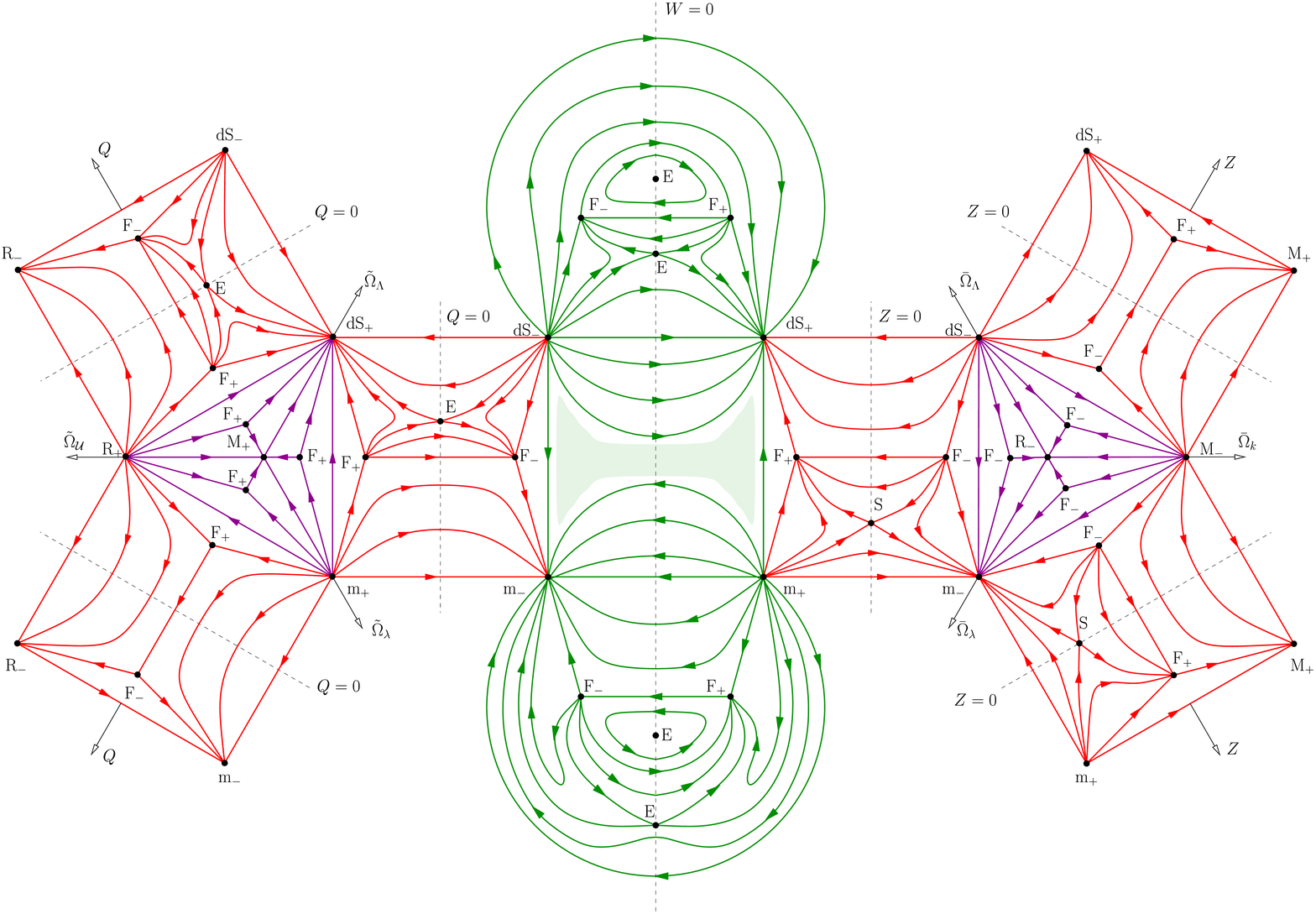}
\caption{Full state space for FLRW models with $\gamma\in
(\textstyle{2\over3},\textstyle{4\over3})$ .}\label{g243}
\end{center}
\end{figure*}

\begin{figure*}
\begin{center}
\includegraphics[height=4in,width=6.5in
]{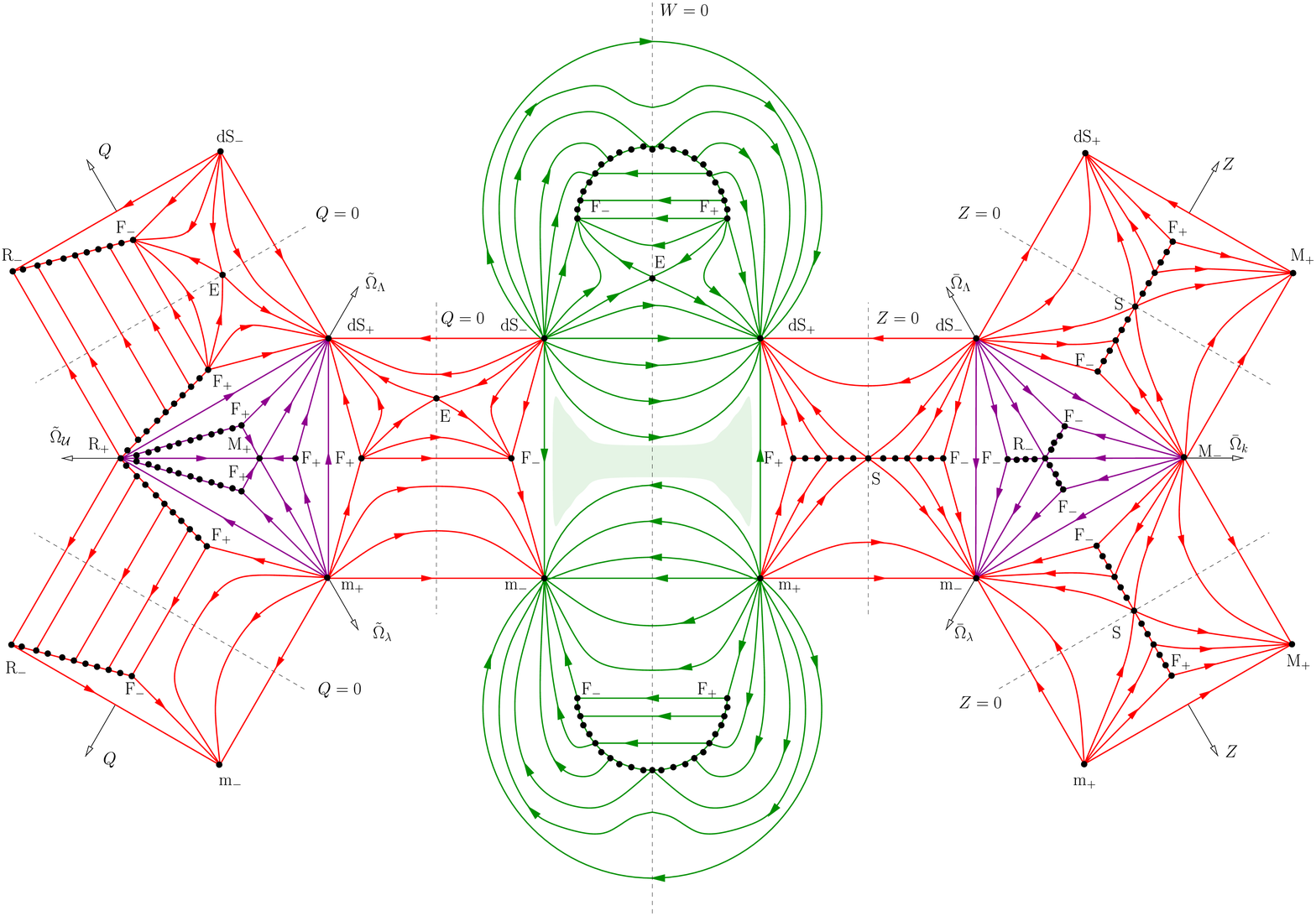}
\caption{Full state space for FLRW models with $\gamma=\textstyle{4\over3}$ .}
\label{g4_3}
\end{center} 
\end{figure*}

\subsection{Qualitative analysis}

Using the information obtained in the previous subsections we are 
going to describe qualitatively the full dynamics of FLRW in brane-world
scenarios.  To that end, we will construct and analyze the structure of
the complete state space for these models.  To start with, we need to
know the dynamical character of the critical points.  This can be
deduced from the eigenvalues given above and summarized in the following
table
\begin{quasitable}
\begin{tabular}{cccc}
Model  & \multicolumn{3}{c}{Dynamical character} \\
\mbox{} &  $0<\gamma < \textstyle{2\over3}$ & $\gamma=\textstyle{2\over3}$ &
$\gamma > \textstyle{2\over3}$  \\ \tableline
$\mbox{F}_{\pm}$ & saddle & saddle & saddle  \\
$\mbox{M}_{\pm}$ & saddle & saddle & saddle  \\
$\mbox{dS}_+$ & attractor & attractor & attractor  \\
$\mbox{dS}_-$ & repeller & repeller & repeller \\
$\mbox{E}~$ & saddle & saddle & saddle        \\
$\mbox{m}_+$  & saddle & repeller & repeller  \\
$\mbox{m}_-$  & saddle & attractor & attractor \\
$\mbox{R}_+$  & repeller & repeller & saddle \\
$\mbox{R}_-$  & attractor & attractor & saddle \\
$\mbox{S}~$ & \mbox{---} & saddle & saddle
\end{tabular}
\end{quasitable}
We can observe some changes with respect to GR and the analysis done
for ${\cal U}=0$ (Paper I).  For instance, the Milne universe is always a 
saddle point for ${\cal U}\neq 0$ whereas $\mbox{M}_+$ ($\mbox{M}_-$) is a 
repeller (attractor) for $\gamma\leq 2/3$ in GR and for $\gamma\leq 1/3$ 
in brane-world scenarios with ${\cal U}=0\,.$ 

The full state space for the FLRW models is four-dimensional, due to the
constraint imposed by the Friedmann equation, and consists of four 
sectors (subsections \ref{frwa}-\ref{frwd}). They are disconnected
in the sense that trajectories cannot change sector.  Here, we will
use a two-dimensional representation of the state space, formed by
all the two-dimensional invariant submanifolds, which in turn determine the
whole dynamics.  We have included the diagrams corresponding
to the cases $\gamma\in(\textstyle{2\over3},\textstyle{4\over3})$ (see
Fig.~\ref{g243}) and $\gamma=\textstyle{4\over3}$ (see Fig.~\ref{g4_3}), 
which contain most of the physically interested cases, like dust ($\gamma=1$)
and radiation ($\gamma=\textstyle{4\over3}$). The second figure constitutes 
a new bifurcation\footnote{The other bifurcations occur at 
$\gamma=0,\textstyle{2\over3}$ (also for GR and ${\cal U}=0$) and 
$\gamma=\textstyle{1\over3}$ (also for ${\cal U}=0$). It is important to 
note that in the case $\gamma=\textstyle{2\over3}$ we find new lines 
of critical points corresponding to models with scale factor $a(t)=t^{1/2}$.} 
due to the structure of the ${\cal U}$-term, which is the same as 
that of a radiation fluid. Then, this bifurcation is characterized by the 
presence of lines of new critical points with scale factor $a(t)=t^{1/2}$ 
but whose dynamical character can be either that of a saddle, repeller 
or attractor, depending on the signs of ${}^3R$ and ${\cal U}$.

\begin{figure}
\begin{center}
\includegraphics[height=3in,width=3in]{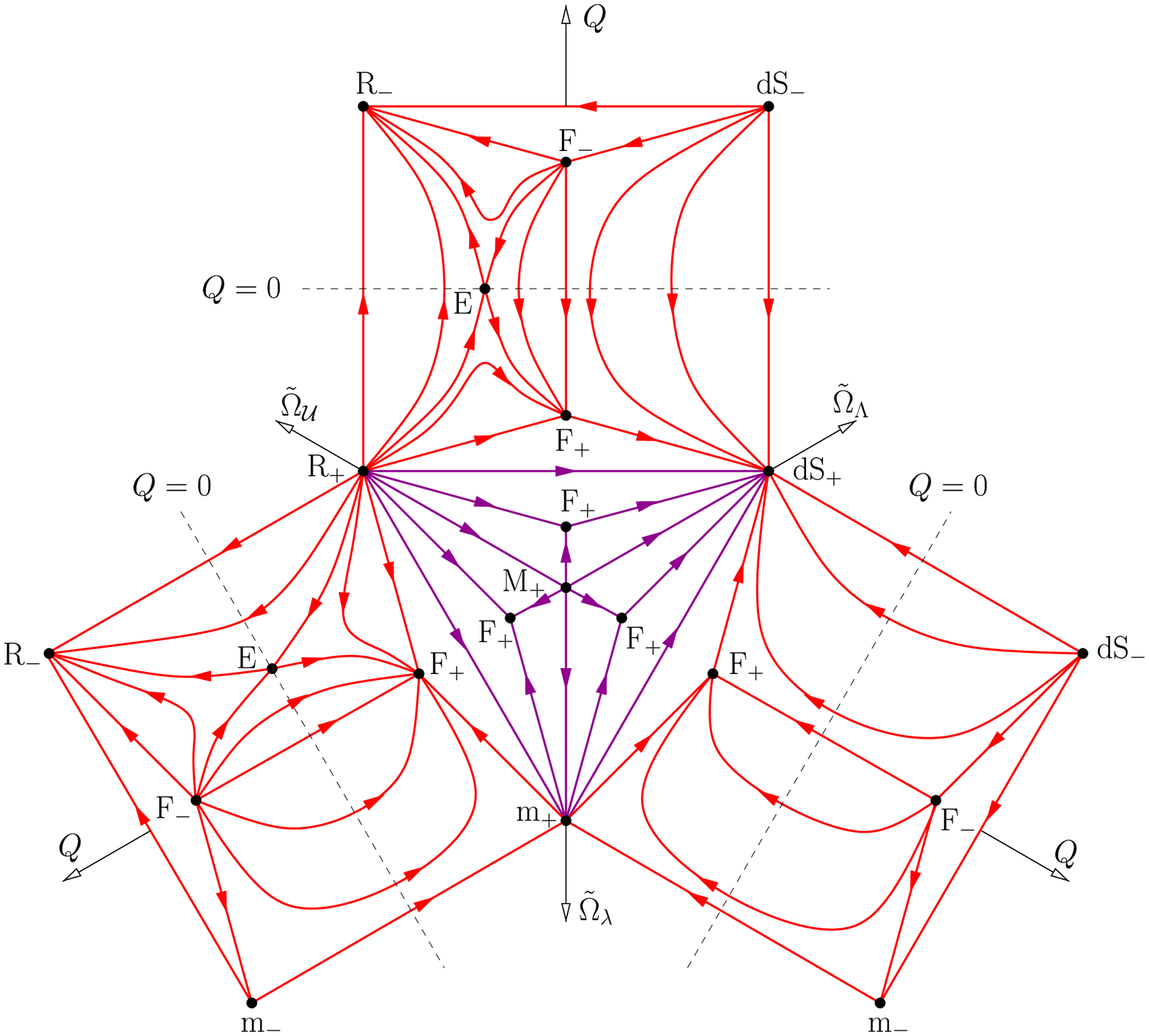}
\caption{Sector corresponding to the dynamics described in 
subsections~\ref{frwa} and \ref{frwb} of the state space for the FLRW 
models with $\gamma\in(0,\textstyle{1\over3})$ .  It shows that 
the Einstein universe appears even for $\gamma \leq \textstyle{1\over3}\,,$  
in contrast to what happens in GR and also when ${\cal U}=0\,.$}\label{g013}
\end{center}
\end{figure}

Coming back to the description of the state space diagrams, the sectors of 
subsections \ref{frwb} and \ref{frwc} are represented by three rectangular 
areas, crossed by the dashed lines $Q=0$ and $Z=0$ respectively, enclosing
triangular areas corresponding to the sector of subsection \ref{frwa}.
The central area, crossed by the line $W=0$, describes the dynamics of the 
sector in subsection \ref{frwd}. The shadowed region does not belong to the
state space, it appears as a result of the two-dimensional representation
we have used.  The dashed lines $Q=Z=W=0$, where the Hubble function vanishes, 
separate regions of expanding models from regions with contracting models.
Using this structure for the state space, the trajectories follow
directly from the information given in the previous subsections.

\begin{figure*}
\begin{center}
\includegraphics[height=3in,width=6.5in
]{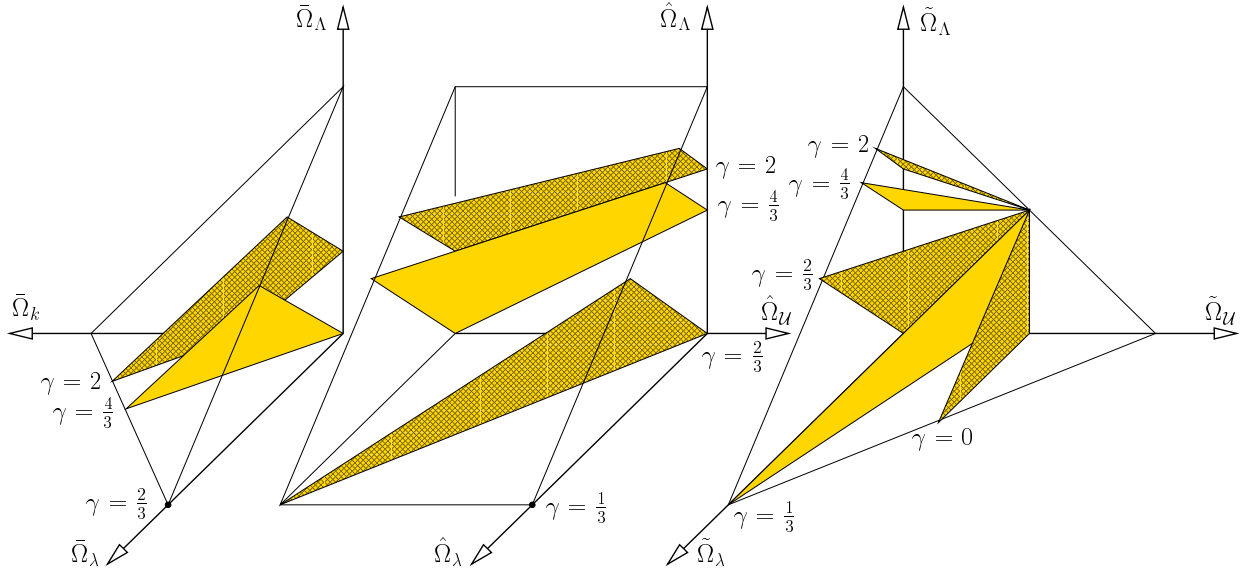}
\caption{Surfaces of saddle critical points corresponding to static FLRW
models ($H=0$). The diagram on the left is the submanifold $Z=0$ of the 
sector ${}^3R\leq 0$ and ${\cal U}\leq 0$  and the 
critical points represented correspond to $\mbox{S}$. The diagrams on the
center (right) show the location of the points $\mbox{E}$ in the 
submanifolds $W=0$ ($Q=0$) of the sector ${}^3R\geq 0$ and ${\cal U}\leq 0$ 
(${}^3R\geq 0$ and ${\cal U}\geq 0$).  Note that $Q=0$, $Z=0$ and $W=0$
are not invariant submanifolds and therefore they do not contain any
trajectories.}\label{sucp}
\end{center}
\end{figure*}

The attractors of the evolution are $\mbox{dS}_+$, $\mbox{R}_-$ and
$\mbox{m}_-$.  More specifically, $\mbox{dS}_+$ is an attractor for any
value of $\gamma$, whereas $\mbox{R}_-$ is an attractor only for
$\gamma\leq 2/3$ and $\mbox{m}_-$ for $\gamma\geq 2/3$. In fact,
$\mbox{R}_-$ and $\mbox{m}_-$ exchange their role as an attractor
at the bifurcation $\gamma=2/3$.  Moreover, while $\mbox{dS}_+$ is an 
attractor for models that at some time will become ever expanding, 
$\mbox{R}_-$ and 
$\mbox{m}_-$ are attractors for (re)collapsing models.  An interesting
feature of the state space is that the volume of the region with models 
evolving towards $\mbox{dS}_+$ decreases as $\gamma$ increases, which
means that the region of models collapsing in the future increases.
The fact of having models that recollapse is directly connected with the
appearance of static models, namely $\mbox{E}$ and $\mbox{S}$, which
are saddle points.  In this sense, it is important to remark that 
the presence of ${\cal U}$ makes the Einstein models to appear for
any value of $\gamma$ (see Fig.~\ref{g013}), and the fact that it can be 
negative leads to the appearance of static models with non-positive spatial 
curvature, i.e. $\mbox{S}$.  Furthermore, $\mbox{E}$ and $\mbox{S}$ are not 
isolated critical points but bidimensional surfaces in the state space.
The location of these surfaces depending on the value of $\gamma$ has
been represented in Fig.~\ref{sucp}. 

\begin{figure*}
\begin{center}
\includegraphics[height=2.4in,width=6.5in
]{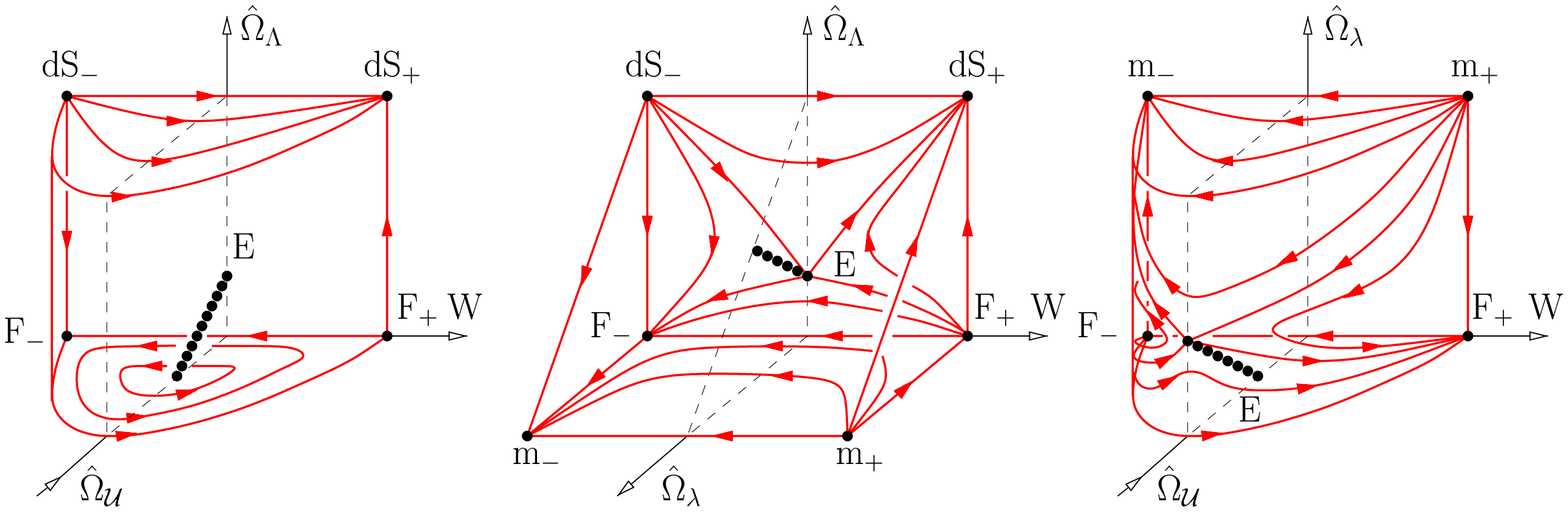}
\caption{Three-dimensional diagrams for the sector ${}^3R\geq 0$ and 
${\cal U}\leq 0$ of the state space of the FLRW models for $\gamma\in
(\textstyle{2\over3},\textstyle{4\over3})$.  Starting from
the left, they correspond to the invariant submanifolds 
$\hat{\Omega}_\lambda=0$, $\hat{\Omega}_{\cal U}=0$ and 
$\hat{\Omega}_\Lambda=0\,.$ The curved part of the boundary of the first 
and third state space diagrams correspond to the invariant submanifold 
$W^2-\hat{\Omega}_{\cal U} =1$.  In these diagrams we also have closed 
trajectories describing oscillating universes.}\label{3ddi}
\end{center}
\end{figure*}

\begin{figure}
\begin{center}
\includegraphics[height=3.3in,width=3.4in]{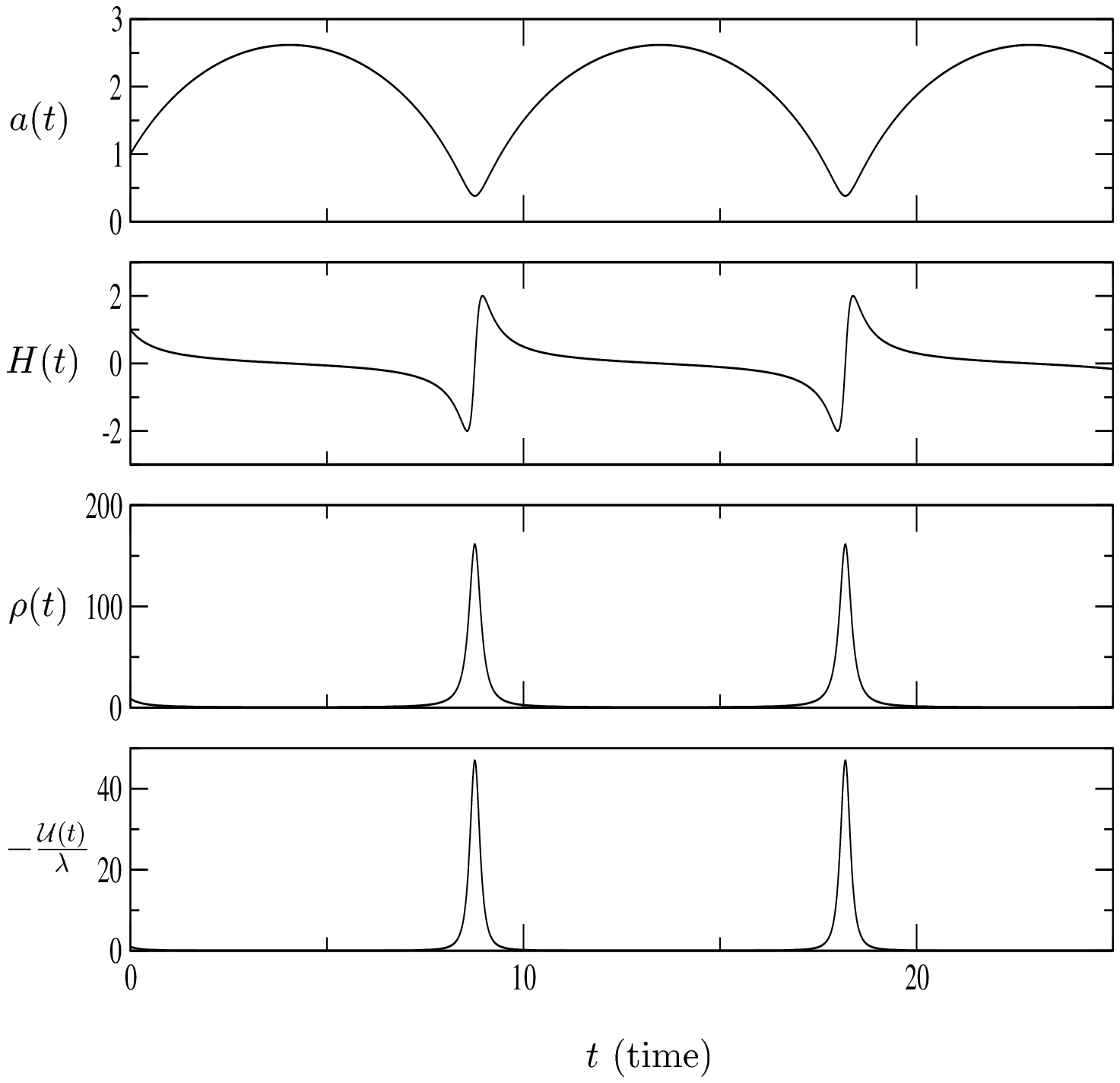}
\caption{Evolution of an oscillating FLRW universe with 
$k=1$, $\gamma=1$ (dust), $\Lambda=0$, and negligible contribution 
from the term quadratic in the energy density.  We have taken
$\kappa=1$ and initial data: $a(0)=H(0)=1$, $\rho(0)=9$ and
${\cal U}(0)=-\lambda$.
}\label{oscu}
\end{center}
\end{figure}

The way of compactifying the state space for the sectors of 
subsections~\ref{frwa} and \ref{frwb} is the same as the one used
in Paper I for the sectors ${}^3R\leq 0$ and ${}^3R\geq 0$ respectively.
The compactification of the sector in subsection~\ref{frwc} is very  
similar to that of subsection~\ref{frwb}, the only difference is that
the negative term in the Friedmann equation is due to ${\cal U}$
instead of ${}^3R$.  In fact, as we can see in Figs.~\ref{g243} and 
\ref{g4_3}, the structure is almost the same.  For the sector 
corresponding to subsection~\ref{frwd} the situation is different due to 
the appearance of two non-positive contributions to the Friedmann equation.  
The compactification of the state space in this case leads to a new type 
of diagrams (see Fig.~\ref{3ddi}).
To understand the dynamics we have to take into account that now we have the 
following  invariant submanifold
\begin{equation} 
W^2-\hat{\Omega}_{\cal U} =1 \,. \label{circ}
\end{equation}
It divides the allowed rectangle $[-1,1]\times [-1,0]$ in the plane 
$(W,\hat{\Omega}_{\cal U})$ into two regions with opposite sign of the
spatial scalar curvature. From the definition of $P$ we find 
\[ W^2-\hat{\Omega}_{\cal U} -1 = \frac{{}^3 R}{6P^2} \,. \]
Since ${}^3R\geq 0$, this means that only the interior region defined
by (\ref{circ}) is allowed (see Fig.~\ref{3ddi}).  The special structure
of this sector of the state space leads to a new exceptional feature
consisting in the appearance of closed trajectories (see Figs.~\ref{g243}
and \ref{3ddi}).  They correspond to cosmological models that after
a certain period of time come back to the same situation as in the 
beginning of that period, and for this reason we will call them 
{\em oscillating universes}.  As we can see in Fig.~\ref{g243}
the trajectories followed by these models have a minimum and a maximum 
value of $W$ which, from Eq.~(\ref{evew}), satisfy the following relation
\[ (1-W^2)\left\{ \frac{3\gamma}{2}(\hat{\Omega}_\rho+
2\hat{\Omega}_\lambda)-1\right\} = -\hat{\Omega}_{\cal U} \,. \]
This equation has solutions only for $\gamma\geq
\textstyle{1\over3}$, just for the values of $\gamma$ that oscillating
universes appear in the state space. Moreover, there is also a minimum and
a maximum of $\hat{\Omega}_{\cal U}$ that corresponds to $W=0$.  
In fact, all the quantities describing these models are periodic and 
therefore have a minimum and a maximum value, in particular the scale 
factor $a(t)$ and the energy density $\rho$.  Hence, these models do not 
have any spacelike singularity.  In Fig.~\ref{oscu} we have represented
the evolution of a particular class of oscillating universes.  Apart
from the features just described it is worth noting that while the
transition from the expanding to the contracting stage is smooth,
the converse is abrupt and mimics a non-singular bang or bounce.
The avoidance of the singularity is due to the negative contribution
of the dark-energy density term ${\cal U}$, which goes like $a^{-4}$, 
to the universe expansion in the Friedmann equation~(\ref{frie}).  In the 
contracting stage this term grows and counterbalances the positive 
contributions in the Friedmann equation, stopping the expansion.  
This can only happen for $\textstyle{1\over3}<\gamma<\textstyle{2\over3}$ 
when the effect of the $\rho^2$ term, which goes like $a^{-6\gamma}$, 
is dominant, and for $\textstyle{2\over3}<\gamma<\textstyle{4\over3}$ 
(see Fig.~\ref{3ddi}) when the effect of the $\rho$ term, which goes 
like $a^{-3\gamma}$, dominates.   After the bounce, the universe expands 
until the spatial curvature ${}^3R$, which goes like $a^{-2}$, stops 
the expansion to enter again a contraction era.

\section{Dynamics of the Bianchi type I models}\label{sec4}
In what follows we will complete the dynamical study of Bianchi type I 
models made in Paper I.  Other works on these models in brane-world
scenarios can be found in~\cite{MASS,VERN,TOPO}.  Exact solutions on the 
brane have been considered in~\cite{CMHM}, and a five-dimensional solution 
for the vacuum case has been presented in~\cite{FROL}.  

As we have discussed above, the non-zero contributions from the 
five-dimensional Weyl tensor are ${\cal U}$ and ${\cal P}_{ab}$ but
since the second one has no evolution equation we will take 
${\cal P}_{ab}=0$.  Then, we have to consider two 
differentiated cases according to the sign of ${\cal U}$: 
(A) ${\cal U}\geq 0$; (B) ${\cal U}\leq 0$.

\subsection{Case ${\cal U}\geq 0$}\label{upos}
The treatment of this case  is very similar to that made in 
subsection~\ref{frwa}.  We will use the dimensionless variables
$\mb{\Omega}=(\Omega_\rho,\Omega_\Lambda,\Omega_\sigma,\Omega_\lambda,
\Omega_{\cal U})$ [See Eqs.(\ref{dlv1}) and (\ref{dlv2})], where we have
defined a normalized variable for the shear contribution 
\begin{equation} 
\Omega_\sigma \equiv \frac{\sigma^2}{3H^2} \,. \label{dlsh}
\end{equation}
The Friedmann equation~(\ref{frie}) then becomes
\[ \Omega_\rho+\Omega_\Lambda+\Omega_\sigma+\Omega_\lambda+
\Omega_{\cal U}=1 \,, \]
and again the state space is compact.  Using the time derivative
$'\equiv |H|^{-1}d/dt$ the system of dynamical equations is
\begin{eqnarray}
\Omega'_\rho & = & \epsilon [2(1+q)-3\gamma]\Omega_\rho \,,
\label{b1i} \\
\Omega'_\Lambda & = & 2\epsilon(1+q)\Omega_\Lambda \,, \\
\Omega'_\sigma & = & 2\epsilon(q-2)\Omega_\sigma \,, \\
\Omega'_\lambda & = & 2\epsilon (1+q-3\gamma)\Omega_\lambda \,, \\
\Omega'_{\cal U} & = & 2\epsilon(q-1)\Omega_{\cal U}\,, \label{b1f}
\end{eqnarray}
where $q$ has the following expression
\[ q = \frac{3\gamma-2}{2}\Omega_\rho+(3\gamma-1)\Omega_\lambda+
2\Omega_\sigma-\Omega_\Lambda+\Omega_{\cal U}\,. \]
The critical points of the dynamical system~(\ref{b1i})-(\ref{b1f}),
their coordinates in the state space and their eigenvalues are given 
in the next table
\begin{quasitable}
\begin{tabular}{ccc}
Model  & Coordinates   & Eigenvalues  \\ \tableline
$\mbox{F}_\epsilon$ & $(1,0,0,0,0)$ & $\epsilon(3\gamma-2,3\gamma,
3(\gamma-2),-3\gamma,3\gamma-4)$ \\
$\mbox{dS}_\epsilon$ & $(0,1,0,0,0)$ & $-\epsilon(3\gamma,2,6,6\gamma,4)$  \\
$\mbox{K}_\epsilon$ & $(0,0,1,0,0)$ & $\epsilon(-3(\gamma-2),6,4,
-6(\gamma-1),2)$ \\ 
$\mbox{m}_\epsilon$ & $(0,0,0,1,0)$ & $2\epsilon(\textstyle{3\gamma\over2},
3\gamma,3(\gamma-1),3\gamma-1,3\gamma-2)$ \\
$\mbox{R}_\epsilon$ & $(0,0,0,0,1)$ & $\epsilon(-(3\gamma-4),4,-2,-2(3\gamma-2),
2)$
\end{tabular}
\end{quasitable}
The critical point $\mbox{K}$ corresponds to the vacuum Kasner models of
GR, whose line element is given by (\ref{lebi}) with  
\begin{equation} 
A_\alpha = t^{2p_\alpha}~~\mbox{and}~~  \sum_{\alpha=1}^3 p_\alpha = 
\sum_{\alpha=1}^3 p^2_\alpha = 1\,. \label{kasn}
\end{equation}
With respect to the analysis of Bianchi type I models done in Paper I, the
only new critical points are the radiation models $\mbox{R}_\pm\,.$

\subsection{Case ${\cal U}\leq 0$}\label{uneg}
The procedure we will follow in this case is the same as in  
subsection~\ref{frwc}.  We will consider the following set of
variables $\mb{\bar{\Omega}}=(Z,\bar{\Omega}_\rho,\bar{\Omega}_\Lambda,
\bar{\Omega}_\sigma,\bar{\Omega}_\lambda)$ [See Eq.~(\ref{deom})],
where $\bar{\Omega}_\sigma$ is the analogous of $\Omega_\sigma$~(\ref{dlsh})
but normalized with respect to $N$~(\ref{deom}).  With this variables
the Friedmann equation~(\ref{frie}) reads
\begin{equation}
\bar{\Omega}_\rho+\bar{\Omega}_\Lambda+\bar{\Omega}_\sigma+
\bar{\Omega}_\lambda = 1 \,, \label{frun}
\end{equation}
and the deceleration parameter $q$ is given by
\[ (q-1)Z^2 = \textstyle{{3\gamma}\over2}(\bar{\Omega}_\rho
+2\bar{\Omega}_\lambda)+3\bar{\Omega}_\sigma-2 \,. \]
With respect to the time derivative $'\equiv N^{-1}d/dt$, the dynamical
system for $\mb{\bar{\Omega}}$ is 
\begin{eqnarray}
Z' & = & -(q-1)Z^2(1-Z^2)\,,  \label{b2i} \\
\bar{\Omega}'_\rho & = & -[(3\gamma-4)-2(q-1)Z^2]Z\bar{\Omega}_\rho \,, \\
\bar{\Omega}'_\Lambda & = & 2[2+(q-1)Z^2]Z\bar{\Omega}_\Lambda\,, \\
\bar{\Omega}'_\sigma & = & -2[1-(q-1)Z^2]Z\bar{\Omega}_\sigma\,, \\
\bar{\Omega}'_\lambda & = & -2[(3\gamma-2)-(q-1)Z^2]Z\bar{\Omega}_\lambda\,.
\label{b2f}
\end{eqnarray}
The critical points of the dynamical system~(\ref{b2i})-(\ref{b2f}),
their coordinates in the state space and their eigenvalues are given 
in the next table~\cite{Nota}
\begin{quasitable}
\begin{tabular}{ccc}
Model  & Coordinates   & Eigenvalues  \\ \tableline
$\mbox{F}_\epsilon$ & $(\epsilon,1,0,0,0)$ & $\epsilon(3\gamma-4,3\gamma,
3\gamma,3(\gamma-2),-3\gamma)$ \\
$\mbox{dS}_\epsilon$ & $(\epsilon,0,1,0,0)$ & $-\epsilon(4,3\gamma,0,6,
6\gamma)$
\\
$\mbox{K}_\epsilon$ & $(\epsilon,0,0,1,0)$ & $\epsilon(2,-3(\gamma-2),6,6,
-6(\gamma-1))$
\\ 
$\mbox{m}_\epsilon$ & $(\epsilon,0,0,0,1)$ & $2\epsilon(3\gamma-2,
\textstyle{{3\gamma}\over2},3\gamma,3(\gamma-1),3\gamma)$ \\
$\mbox{nE}$ & $(0,\bar{\Omega}^\ast_\rho,\bar{\Omega}^\ast_\Lambda,
\bar{\Omega}^\ast_\sigma,\bar{\Omega}^\ast_\lambda)$ & 
$(0,\sqrt{\zeta},0,0,-\sqrt{\zeta})$
\end{tabular}
\end{quasitable}
We have a new set of saddle critical points, designated by
$\mbox{nE}$, whose coordinates in the state space $\mb{\bar{\Omega}^\ast}$
satisfy~(\ref{frun}) and
\[ \textstyle{{3\gamma}\over2}(\bar{\Omega}^\ast_\rho
+2\bar{\Omega}^\ast_\lambda)+3\bar{\Omega}^\ast_\sigma-2 =0\,.\]
The quantity $\zeta$ is given by
\[ \zeta = \frac{3\gamma}{2}\left[(3\gamma-4)\bar{\Omega}^\ast_\rho
+4(3\gamma-2)\bar{\Omega}^\ast_\lambda\right]+6\bar{\Omega}^\ast_\sigma\,, \]
and in the FLRW limit, $\sigma=0\,,$ it corresponds to the quantity
$\psi$ in~(\ref{psie}) for $k=0$.
The $\mbox{nE}$ models are non-expanding ($H=0$) Bianchi models which
in the FLRW limit become static, in fact, in this limit, they coincide with 
the critical points $\mbox{S}$ with ${}^3R=0$. 
A noteworthy difference is that the models $\mbox{S}$ only appear
for $\gamma\geq 2/3$ whereas the models $\mbox{nE}$ appear for all 
values of $\gamma$.
Moreover, $\rho\,,$ $\sigma$ and ${\cal U}$ 
are constants, $(\rho^\ast,\sigma^\ast,{\cal U}^\ast)\,,$
which, in order to have a positive energy 
density must satisfy a condition analogous to~(\ref{cped})
\[ -{\cal U}^\ast \geq \frac{\lambda\kappa^2}{6}\left(\Lambda+
{\sigma^\ast}^2\right) \,. \]
In fact, integrating the effective  Einstein equations~(\ref{mefe}), we can 
find the metric functions $A_\alpha(t)$ in~(\ref{lebi}) corresponding 
to the critical points $\mbox{nE}$,   
\[ A_\alpha(t) = \mbox{e}^{q_\alpha t}\,, \]
where $q_\alpha$ ($\alpha=1-3$) are constants (proportional to the components 
of the shear tensor) such that
\[ \sum_{\alpha =1}^3 q_\alpha = 0 ~~\mbox{and}~~   
   \sum_{\alpha =1}^3 q^2_\alpha = 2{\sigma^\ast}^2\,.\]
Note that despite the scalar factor remains constant ($\prod_{\alpha=1}^3
A_\alpha(t)=1$) the matter can expand or contract along the shear principal 
axes.  In fact, when only one $q_\alpha$ is negative the system evolves 
towards a {\em pancake} singularity, whereas when two of them are negative it 
evolves towards a {\em cigar} singularity (for details on this type of 
singularities see, e.g.,~\cite{STEP}).

\subsection{Qualitative analysis}

To analyze the dynamics we have to consider first the character of the
critical points, which can be deduced from the information given in the
previous subsections and can be summarized in the following table
\begin{quasitable}
\begin{tabular}{cccc}
Model  & \multicolumn{3}{c}{Dynamical character} \\
\mbox{} &  $0<\gamma < 1$ & $\gamma=1$ & $\gamma > 1$  \\ \tableline
$\mbox{F}_{\pm}$ & saddle & saddle & saddle  \\
$\mbox{dS}_+$ & attractor & attractor & attractor  \\
$\mbox{dS}_-$ & repeller & repeller & repeller \\
$\mbox{K}_+$  & repeller & repeller & saddle \\
$\mbox{K}_-$  & attractor & attractor & saddle \\
$\mbox{m}_+$  & saddle & repeller & repeller  \\
$\mbox{m}_-$  & saddle & attractor & attractor \\
$\mbox{R}_{\pm}$ & saddle & saddle & saddle \\
$\mbox{nE}$ & saddle & saddle & saddle
\end{tabular}
\end{quasitable}
The character of the critical points $\mbox{F}_\pm$, $\mbox{dS}_\pm$,
$\mbox{K}_\pm$, and $\mbox{m}_\pm$ is the same as in the case ${\cal U}=0$
studied in Paper I.  The new dynamical features are due to the appearance 
of the critical points $\mbox{R}_\pm$ and $\mbox{nE}$.  Moreover, we also 
find two new bifurcations at $\gamma=\textstyle{2\over3}$ and 
$\gamma=\textstyle{4\over3}$, besides the bifurcations at
$\gamma=0,1,2$ already found in Paper I for the case ${\cal U}=0$.

\begin{figure}
\begin{center}
\includegraphics[height=3in,width=3in]{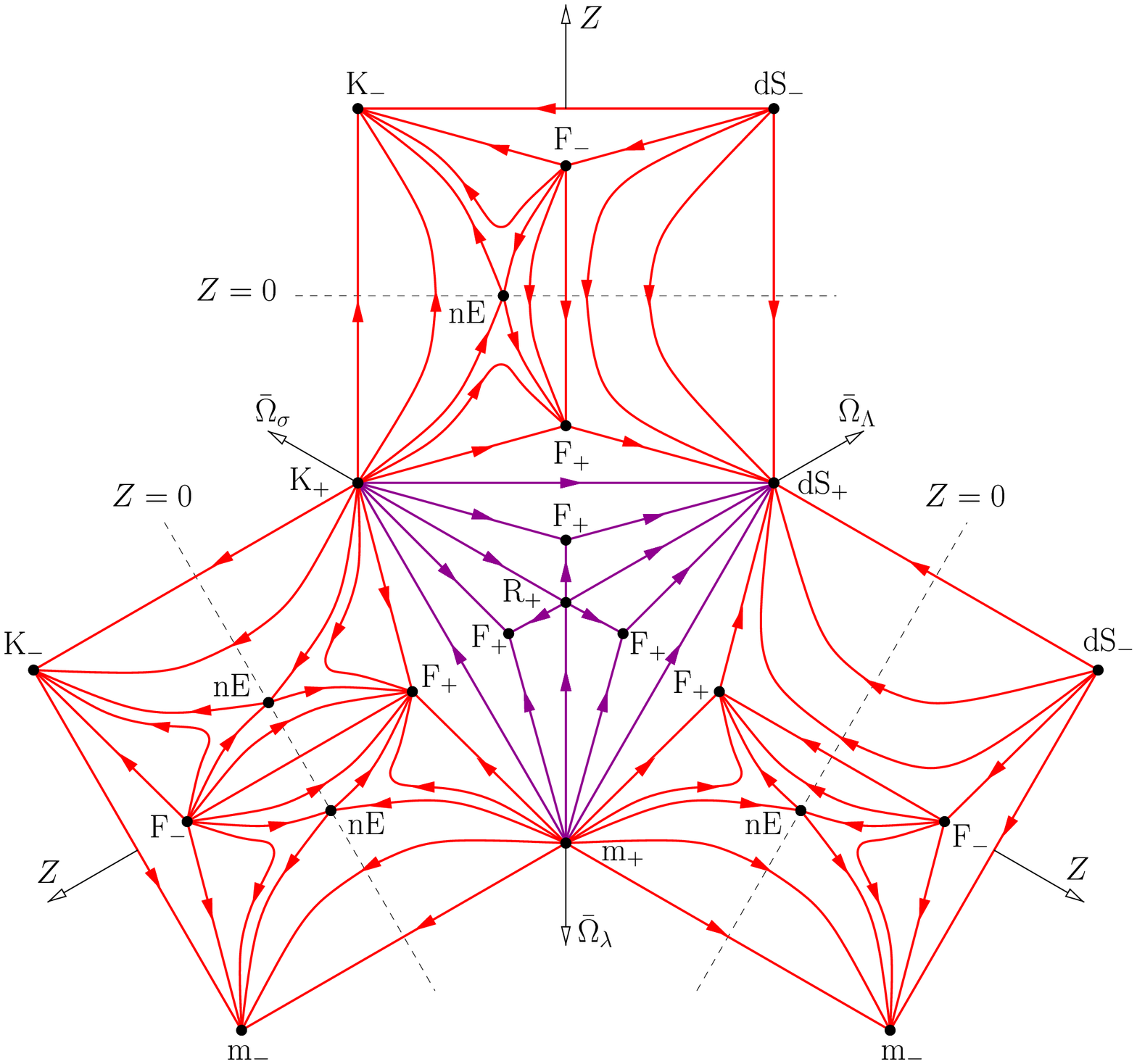}
\caption{Complete state space for Bianchi type I models with 
$\gamma\in(1,\textstyle{4\over3})\,.$}\label{g143}
\end{center}
\end{figure}

\begin{figure}
\begin{center}
\includegraphics[height=3in,width=3in]{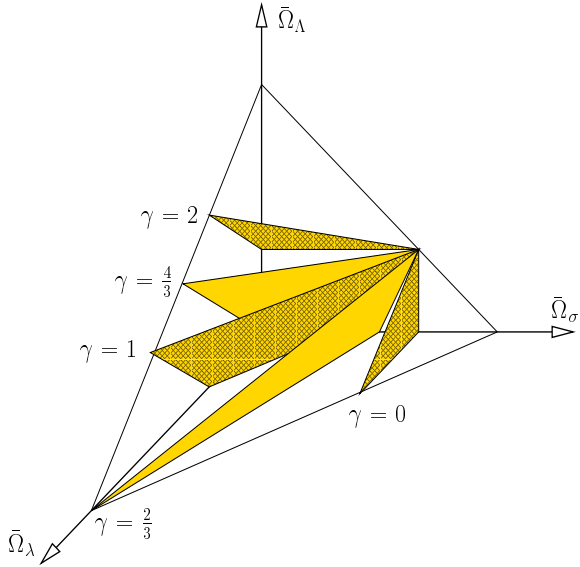}
\caption{Representation of surfaces of saddle critical points corresponding 
to the non-expanding models $\mbox{nE}$ in the $Z=0$ submanifold
of the state space of Bianchi type I models.}\label{cpne}
\end{center}
\end{figure}

The state space for Bianchi type I models is also four-dimensional and is
divided into two sectors according to the sign of ${\cal U}$.  We have used
the same type of bidimensional representation in which only the trajectories
for the two-dimensional invariant submanifolds are drawn.  In Fig.~\ref{g143} 
we show the state space for $\gamma\in(1,\textstyle{4\over3})$, which 
includes physically well-motivated situations from dust to radiation 
universes.  The sector ${\cal U}\leq 0$ described in subsection~\ref{uneg} 
is represented by three rectangular regions which enclose a triangle
where the dynamics of the sector ${\cal U}\geq 0$ (subsection~\ref{upos})
is depicted. The dashed lines $Z=0$ divided regions of expanding
models from regions of contracting models, and these are the places
where the non-expanding models $\mbox{nE}$ appear.  In fact, these saddle 
points are not isolated points but two-dimensional surfaces.  The location 
of the surfaces of points $\mbox{nE}$ for different values of $\gamma$ has 
been drawn in Fig.~\ref{cpne}.

For the sector ${\cal U}\geq 0$ the general attractor is de Sitter,
as it happens when ${\cal U}=0$.  The situation changes for the
sector ${\cal U}\leq 0$, where the possible final states of the evolution 
are $\mbox{dS}_+$, $\mbox{K}_-$ and $\mbox{m}_-$.  Like in some sectors
of state space of the FLRW models, the sector ${\cal U}\leq 0$ of Bianchi
models is divided into two regions, the region of models that will approach 
$\mbox{dS}_+$ and then will become ever expanding, and the region of models
that will (re)collapse.  Obviously the models approaching $\mbox{dS}_+$
in the future isotropize, although there is an intermediate stage in
which the relative contribution of shear grows and reaches a maximum.  
In the sector ${\cal U}\geq 0$ this maximum is reached 
when the following condition holds (see~\cite{TOPO})
\[ \Omega_\sigma = 1-\textstyle{2\over3}\Omega_{\cal U}
-\textstyle{\gamma\over2}(\Omega_\rho+2\Omega_\lambda) \,, \]
and in the sector ${\cal U}\leq 0$ the condition is
\[ \bar{\Omega}_\sigma = 1-\textstyle{\gamma\over2}(\bar{\Omega}_\rho+
2\bar{\Omega}_\lambda) \,. \]
On the other hand, the attractor for collapsing models (${\cal U}\leq 0$) is 
in general $\mbox{K}_-$  ($\mbox{m}_-$ for the submanifold $\sigma=0$)
for $0<\gamma< 1$ and $\mbox{m}_-$ ($\mbox{K}_-$ for the submanifolds
$\bar{\Omega}_\lambda=0$) for $\gamma> 1$.  For dust
($\gamma=1$) we have a bifurcacion in which the attractors are anisotropic 
models whose line element is the same as the one of Kasner models [see
Eqs.(\ref{lebi}),(\ref{kasn})] but the constant $p_\alpha$ now only satisfy
\[ \sum_{\alpha=1}^3 p_\alpha =1 \,. \]
From this discussion we deduce that there are sets of trajectories in the 
state space that either do not isotropize in the future or isotropize but 
the final state is not de Sitter ($\mbox{dS}_+$).  Hence, the 
cosmic no-hair theorem is not satisfied in the ${\cal U}\leq 0$ sector.
This fact has been anticipated in the analysis made in~\cite{VERN}, 
although in their non-compact representation of the state space the final 
state for collapsing models was not shown.  

Finally, with regard to the structure of the initial singularity, in 
Paper I we showed that for ${\cal U}=0$ and $\gamma\geq 1$ the initial 
singularity is isotropic, in contrast to what happens in GR.  Now, this is 
also true for the sector ${\cal U}> 0$.  In the sector ${\cal U}< 0$ 
the singularity is also isotropic except for the submanifold 
$\bar{\Omega}_\lambda=0$.

\section{Remarks and conclusions}\label{core}

In this paper we have completed the dynamical systems analysis 
on the evolution of cosmological models in brane-world type scenarios
started in Paper I, where the effects of the five-dimensional (bulk) 
Weyl tensor were not considered.  By including this quantity in the
analysis we have constructed the most complete state space for 
non-tilted perfect-fluid FLRW cosmological models.  In the case of 
Bianchi type I models the state space we have
presented is also complete except for the fact that we have disregarded 
the components of the five-dimensional Weyl tensor for which the theory
does not provide evolution equations, namely ${\cal P}_{ab}$.
This means that we cannot make predictions on how the term ${\cal P}_{ab}$ 
affects the dynamics.  Perhaps future studies on string/M-theory could shed 
light on this question.

From the dynamical study carried out we have found new interesting
features not present in GR nor in brane-world scenarios with ${\cal U}=0$.
In particular, we have discovered new critical points: $\mbox{R}_\pm$,
$\mbox{S}$ and $\mbox{nE}$.  The Einstein universe $\mbox{E}$ and the 
non-expanding anisotropic models $\mbox{nE}$ appear for any value
of $\gamma$, which is directly connected to the fact that we have
now (re)collapsing models both for FLRW and Bianchi type I models for any
$\gamma$.   It is also worth noting that now we have static critical
points for any sign of the spatial curvature ${}^3R$.  We have also
seen that de Sitter ($\mbox{dS}_+$) is always an attractor but not
the only one.  For FLRW models, $\mbox{R}_-$ and $\mbox{m}_-$ are also 
attractors for $\gamma \leq \textstyle{2\over3}$ and 
$\gamma \geq \textstyle{2\over3}$ respectively, and for Bianchi type
I models, $\mbox{K}_-$ and $\mbox{m}_-$ are attractors for $\gamma\leq 1$
and $\gamma\geq 1$ respectively.

A relevant interesting feature
is the appearance of oscillating universes in the ${}^3R\geq 0$
and ${\cal U}\leq 0$ sector of the state space of the FLRW models.
In these universes the physical variables oscillate periodically between
a minimum and a maximum value without reaching any spacelike singularity.   
With regard to the evolution of the anisotropy in Bianchi type I models we 
have found that for ${\cal U}\geq 0$  they always isotropize whereas for 
${\cal U}\leq 0$ they can both isotropize or collapse.  
This violation of the cosmic no-hair theorem has also been pointed out 
in~\cite{VERN}.  
The use in the present paper of dynamical variables that compactify the
state space allows us to see clearly which are the possible final states 
of the collapse ($\mbox{K}_-$ or $\mbox{m}_-$) and
how the state space is divided into two regions according to whether
the models isotropize or collapse.  On the other hand, we have seen
that the initial singularity is isotropic for $\gamma\geq 1$, as it
also happens in the particular case ${\cal U}=0$.

Finally, the analysis we have carried out in this paper
provides information on the question of the stability of the so-called 
Randall-Sundrum fine-tuning condition, which consists in the vanishing 
of the four-dimensional cosmological constant ($\Lambda=0$).   
More specifically, in brane-world scenarios the four-dimensional 
cosmological constant is given by
\[ \Lambda = \frac{|\Lambda_{(5)}|}{2}\;\left[\left(
\frac{\lambda}{\lambda_c}\right)^2-1\right] \,, ~~~
\lambda^2_c\equiv 6\frac{|\Lambda_{(5)}|}{\kappa^4_{(5)}}\,,\]
where $\lambda_c$ is a critical tension.  Then the fine tuning consists
in adjusting the brane tension $\lambda$ so that $\lambda=\lambda_c$.  
From our perspective, we can deal with this question by asking whether or
not  the invariant submanifold $\Omega_\Lambda=0$ (and also the equivalent
ones $\tilde{\Omega}_\Lambda=\bar{\Omega}_\Lambda=\hat{\Omega}_\Lambda=0$), 
where the fine-tuning condition holds, is dynamically stable.  In other 
words, what happens when initial conditions are prescribed near 
this invariant submanifold of the state space.  It can
be seen that for the regions of recollapsing models the evolution
will bring them back to $\Omega_\Lambda=0$, although if we start with
$H>0$ the models will initially separate from that submanifold.
For the other regions, the evolution will move the models away from 
$\Omega_\Lambda=0$ in such a way that they will approach de Sitter, 
which is precisely the opposite situation to $\Lambda=0$ (see also~\cite{CARS}
for an alternative discussion).  Then, we can conclude that the fine-tuning 
condition is not dynamically stable in the sense 
that there are non-zero measure regions of initial data {\em near} the
fine-tuning $\Lambda=0$ evolving towards a situation in which that condition
does not longer hold.



\[ \]
{\bf Acknowledgements:}
The authors wish to thank their colleagues in the Relativity and Cosmology
Group for fruitful discussions.  This work has been supported by the
European Commission (contracts HPMF-CT-1999-00149 and HPMF-CT-1999-00158).





\begin{thebibliography}{10}
\bibitem{Nota} The presence of a zero eigenvalue is (excepting for  
bifurcations and the sets of critical points $\mbox{E}$, $\mbox{S}$ and
$\mbox{nE}$) a consequence of having
a dynamical system with a constraint: The Friedmann equation. If we had
used it to eliminate one variable and reduce the dimension of the state 
space, we would not find any vanishing eigenvalue.
In conclusion, this fact does not affect to the hyperbolic character of 
the critical points.
\end{thebibliography}
\end{document}